\documentclass[letterpaper]{article} 
\usepackage{aaai24}  
\usepackage{times}  
\usepackage{helvet}  
\usepackage{courier}  
\usepackage[hyphens]{url}  
\usepackage{graphicx} 
\usepackage{natbib}  
\usepackage{caption} 
\usepackage{algorithm}
\usepackage{algorithmic}

\usepackage{subfigure}
\usepackage{amsmath,amssymb}
\usepackage{multirow}
\usepackage{newfloat}
\usepackage{listings}
\DeclareCaptionStyle{ruled}{labelfont=normalfont,labelsep=colon,strut=off} 
\floatstyle{ruled}
\newfloat{listing}{tb}{lst}{}
\floatname{listing}{Listing}
\title{UniCATS: A Unified Context-Aware Text-to-Speech Framework\\ with Contextual VQ-Diffusion and Vocoding}
\author{
    Chenpeng Du\textsuperscript{\rm 1},
    Yiwei Guo\textsuperscript{\rm 1},
    Feiyu Shen\textsuperscript{\rm 1},
    Zhijun Liu\textsuperscript{\rm 1},
    Zheng Liang\textsuperscript{\rm 1},\\
    Xie Chen\textsuperscript{\rm 1},
    Shuai Wang\textsuperscript{\rm 2},
    Hui Zhang\textsuperscript{\rm 3},
    Kai Yu\textsuperscript{\rm 1}\thanks{Corresponding author.}
}
\affiliations{
    \textsuperscript{\rm 1}MoE Key Lab of Artificial Intelligence, AI Institute \\
    X-LANCE Lab, Department of Computer Science and Engineering, Shanghai Jiao Tong University, Shanghai, China \\
    \textsuperscript{\rm 2} Shenzhen Research Institute of Big Data, Shenzhen, China \\
    \textsuperscript{\rm 3} AISpeech Ltd, Beijing, China \\
    \{duchenpeng, kai.yu\}@sjtu.edu.cn
}

\usepackage{bibentry}

\begin{document}

\maketitle

\begin{abstract}

The utilization of discrete speech tokens, divided into semantic tokens and acoustic tokens, has been proven superior to traditional acoustic feature mel-spectrograms in terms of naturalness and robustness for text-to-speech (TTS) synthesis. Recent popular models, such as VALL-E and SPEAR-TTS, allow zero-shot speaker adaptation through auto-regressive (AR) continuation of acoustic tokens extracted from a short speech prompt. 
However, these AR models are restricted to generate speech only in a left-to-right direction, making them unsuitable for speech editing where both preceding and following contexts are provided. Furthermore, these models rely on acoustic tokens, which have audio quality limitations imposed by the performance of audio codec models.
In this study, we propose a unified context-aware TTS framework called UniCATS, which is capable of both speech continuation and editing. UniCATS comprises two components, an acoustic model CTX-txt2vec and a vocoder CTX-vec2wav. 
CTX-txt2vec employs contextual VQ-diffusion to predict semantic tokens from the input text, enabling it to incorporate the semantic context and maintain seamless concatenation with the surrounding context. Following that, CTX-vec2wav utilizes contextual vocoding to convert these semantic tokens into waveforms, taking into consideration the acoustic context.
Our experimental results demonstrate that CTX-vec2wav outperforms HifiGAN and AudioLM in terms of speech resynthesis from semantic tokens. Moreover, we show that UniCATS achieves state-of-the-art performance in both speech continuation and editing. Audio samples are available at https://cpdu.github.io/unicats.

\end{abstract}

\section{Introduction}

Recently, two types of discrete speech tokens have been proposed, which are known as semantic tokens and acoustic tokens \cite{audiolm}. Semantic tokens, such as vq-wav2vec \cite{vqw2v}, wav2vec 2.0 \cite{w2v2} and HuBERT \cite{hubert}, are trained for discrimination or masking prediction. Consequently, they primarily capture articulation information while providing limited acoustic details. On the other hand, acoustic tokens, which have been introduced by audio codec models like Soundstream \cite{soundstream} and Encodec \cite{encodec}, are trained specifically for speech reconstruction. As a result, they capture acoustic details, especially speaker identity.

\begin{figure}[t]
\centering
\includegraphics[page=1,width=0.98\linewidth,trim=7cm 3.4cm 4.8cm 0cm,clip=true]{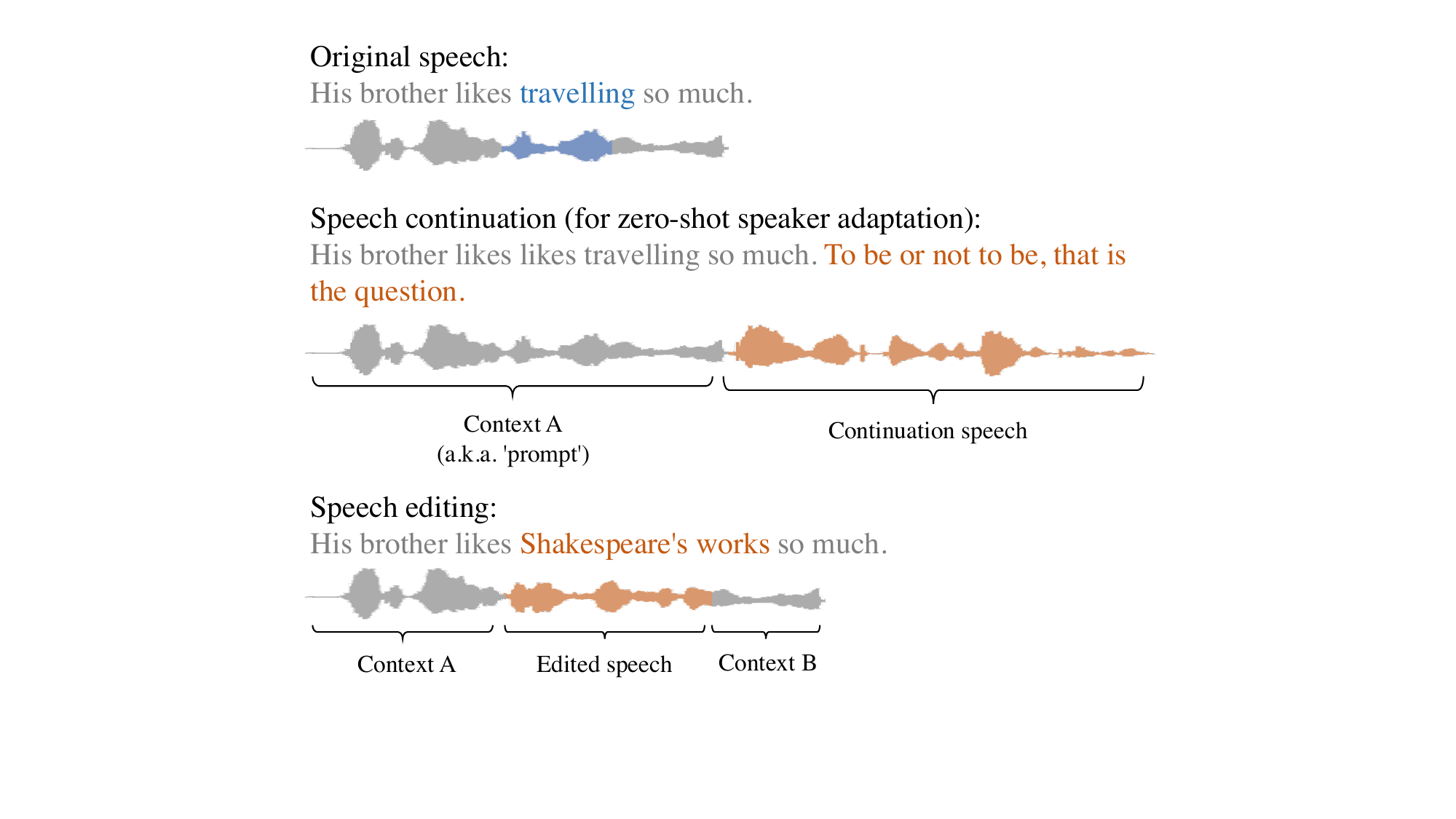}
\caption{Definitions of context-aware TTS tasks, including speech continuation and speech editing.}
\label{fig:task}
\end{figure}

\begin{figure*}[t]
  \centering
  \subfigure[Speech continuation]{
    \includegraphics[page=3,width=0.48\linewidth,trim=0.8cm 0.5cm 11.6cm 1cm,clip=true]{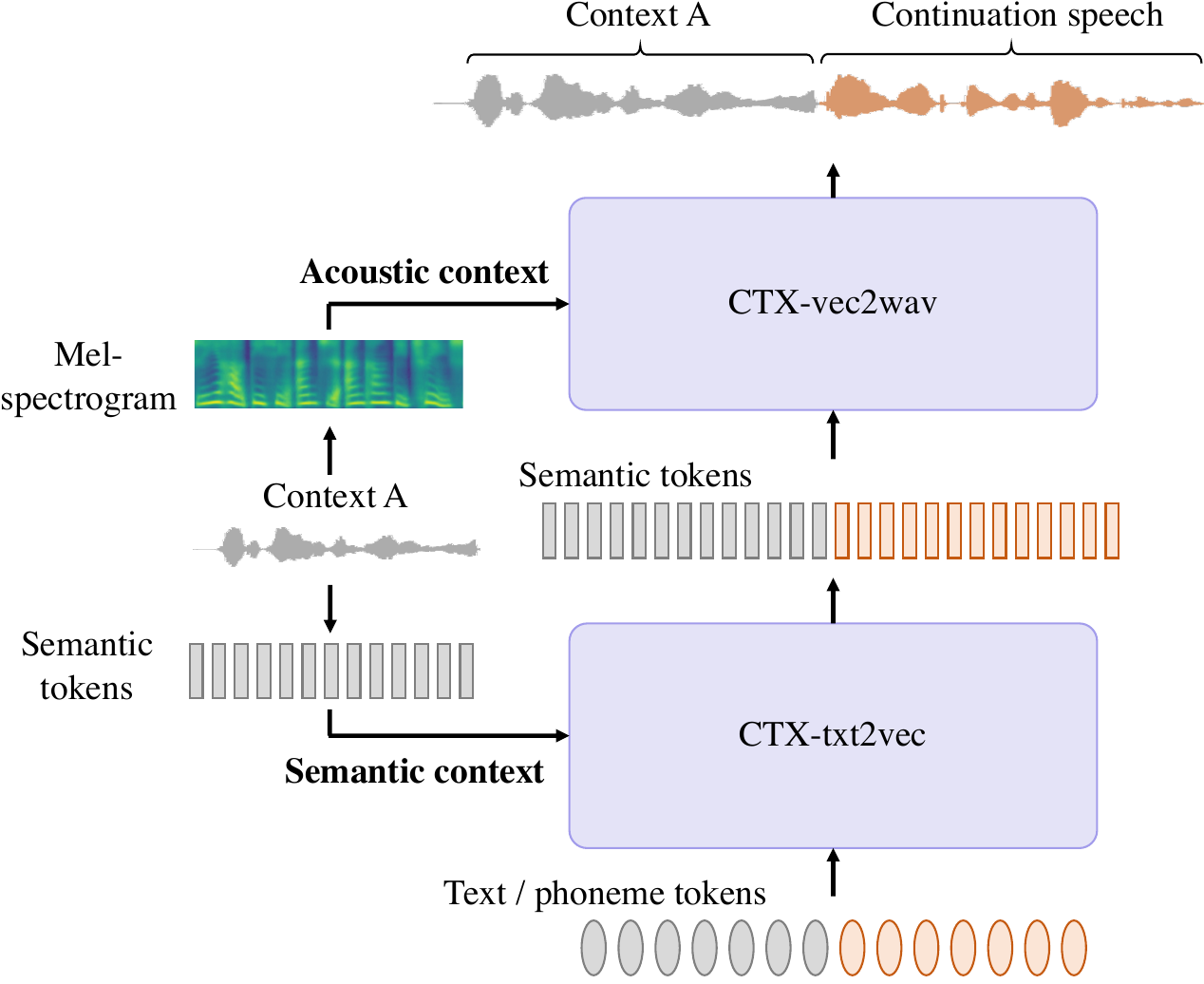}
    \label{pipeline_continuation}
  }
  \subfigure[Speech editing]{
    \includegraphics[page=2,width=0.48\linewidth,trim=1.6cm 0.5cm 10.8cm 1cm,clip=true]{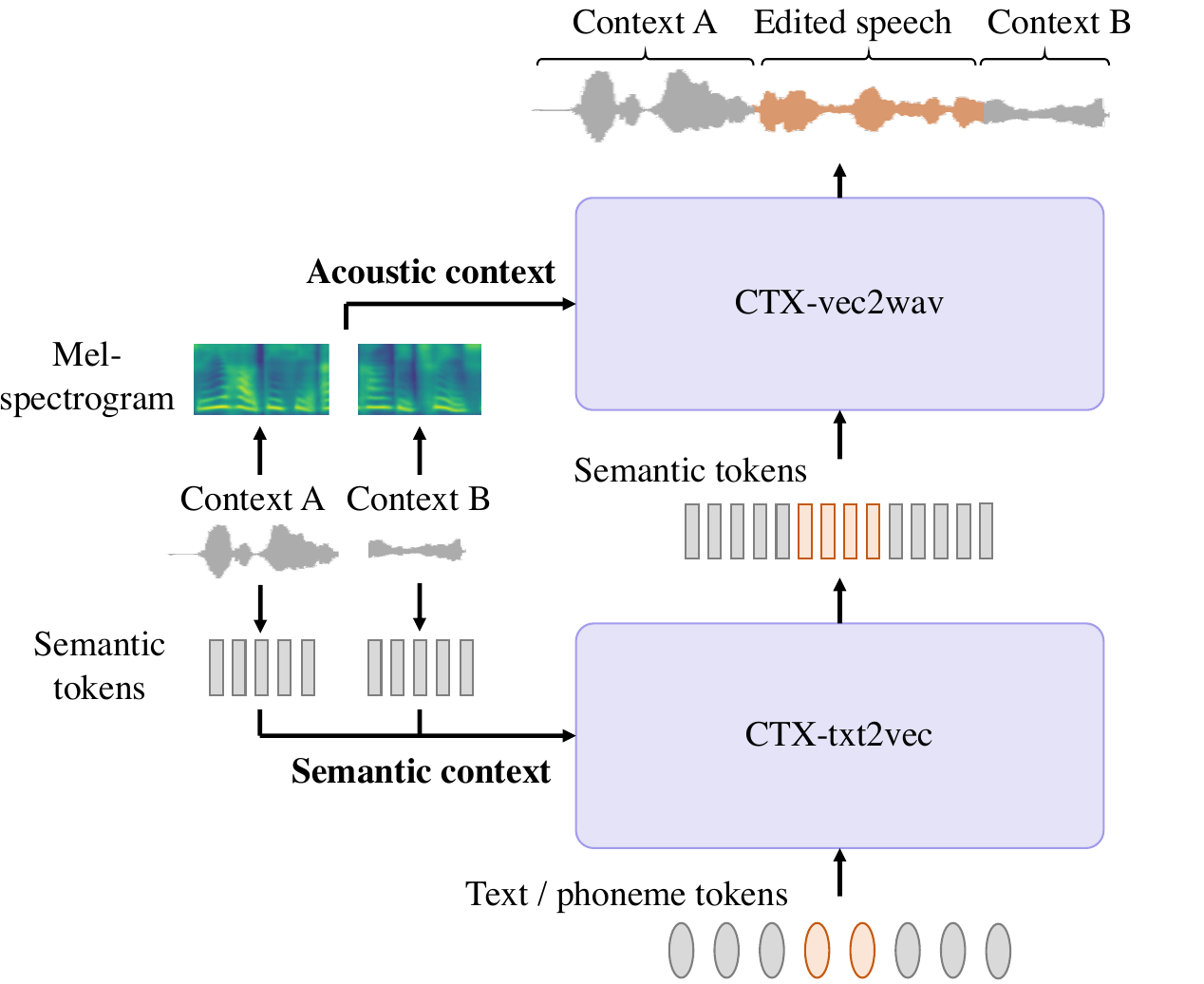}
    \label{pipeline_editing}
  }
  \caption{The unified context-aware framework UniCATS for speech continuation and editing. Both the two tasks share the same model, with the only distinction being the presence of context B.}
  \label{fig:UniCATS}
\end{figure*}

The typical neural text-to-speech (TTS) pipeline, such as Tacotron 2 \cite{tacotron2} and FastSpeech 2 \cite{fastspeech2}, consists of two stages: predicting the mel-spectrogram from text and then vocoding it into waveform. Additional techniques, such as normalizing flow \cite{flowtron,glowtts} and diffusion models \cite{diffsinger,gradtts,diffvoice}, have been introduced to generate the mel-spectrogram. Recently, VQTTS \cite{vqtts} proposes a novel approach by utilizing discrete speech tokens as the intermediate representation for text-to-speech synthesis. The discrete tokens have been found to exhibit superior naturalness and robustness compared to mel-spectrograms. Textless NLP \cite{testlessnlp} and AudioLM \cite{audiolm} propose to leverage wav2vec 2.0 and w2v-BERT \cite{w2vbert} respectively for language model training and consequently are able to generate speech unconditionally via auto-regressive inference. InstructTTS \cite{instructtts} uses VQ-diffusion to generate acoustic tokens whose speaking style is guided by a natural language prompt.
 VALL-E \cite{valle} and SPEAR-TTS \cite{speartts} further extend the use of discrete tokens to zero-shot speaker adaptation. Specifically, they generate acoustic tokens based on the input text using a decoder-only auto-regressive (AR) model. During inference, they conduct AR continuation from the acoustic tokens of a short speech prompt provided by the target speaker. As a result, these models are capable of generating speech in the target speaker's voice. 
NaturalSpeech 2 \cite{naturalspeech2} employs a typical diffusion model to generate discrete acoustic tokens as continuous features.

In addition to speech continuation, there is another context-aware TTS task called speech editing \cite{editts,retrievertts}. Illustrated in Figure \ref{fig:task}, speech editing means synthesizing speech based on input text while ensuring smooth concatenation with its surrounding context. Unlike speech continuation, speech editing takes into account both the preceding context A and the following context B.

However, current TTS models that based on discrete speech tokens face three limitations. Firstly, most of these models are autoregressive (AR) models, which restricts them to generate speech only in a left-to-right direction. This limitation makes them unsuitable for speech editing, where both preceding and following contexts are provided. Secondly, the construction of acoustic tokens involves residual vector quantization (RVQ), resulting in multiple indices for each frame. This approach introduces prediction challenges and complexity into text-to-speech. For instance, VALL-E incorporates a non-auto-regressive (NAR) module to generate the residual indices, while SPEAR-TTS addresses this issue by padding the RVQ indices into a longer sequence, which further complicates modeling. Lastly, the audio quality of these TTS systems is constrained by the performance of audio codec models.

\begin{figure*}[t]
  \centering
  \subfigure[Overall architecture]{
    \includegraphics[page=4,width=0.57\linewidth,trim=0.8cm 2cm 10cm 4cm,clip=true]{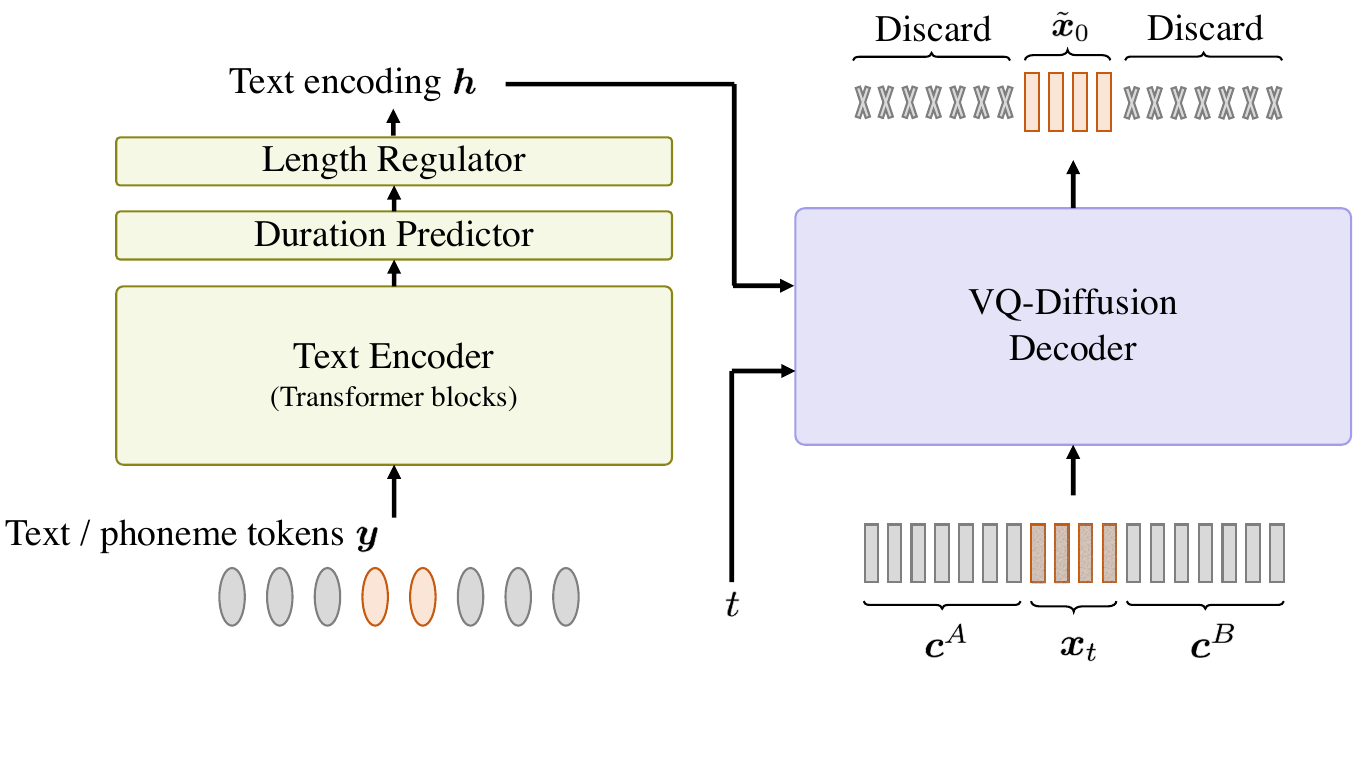}
    \label{ict2v_overall}
  }
  \subfigure[VQ-diffusion decoder]{
    \includegraphics[page=5,width=0.39\linewidth,trim=7.4cm 1.5cm 9.5cm 0.8cm,clip=true]{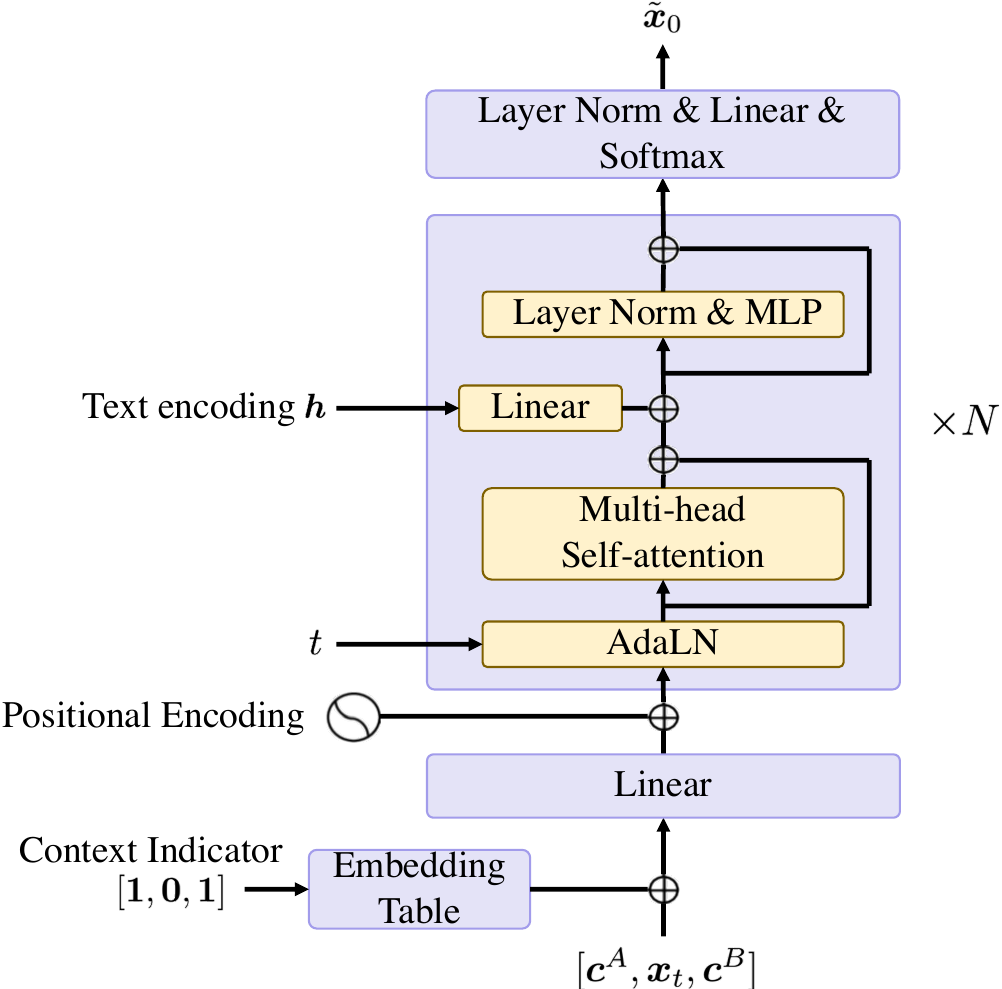}
    \label{vqdiff}
  }
  \caption{The model architecture of CTX-txt2vec with contextual VQ-diffusion.}
  \label{fig:ict2v}
\end{figure*}

To tackle all the three limitations, we propose a unified context-aware TTS framework called UniCATS in this study, designed to handle both speech continuation and editing tasks. UniCATS comprises two components: an acoustic model CTX-txt2vec and a vocoder CTX-vec2wav. Figure \ref{fig:UniCATS} illustrates the pipelines of UniCATS for speech continuation and editing. In these pipelines, CTX-txt2vec employs contextual VQ-diffusion to predict semantic tokens from the input text, enabling it to incorporate the semantic context and maintain seamless concatenation with the surrounding context. Following that, CTX-vec2wav utilizes contextual vocoding to convert these semantic tokens into waveforms, taking into consideration the acoustic context, especially speaker identity. Both speech continuation and editing tasks in UniCATS share the same model, with the only distinction being the presence of context B.
Our experiments conducted on the LibriTTS dataset \cite{libritts} demonstrate that CTX-vec2wav outperforms HifiGAN and AudioLM in terms of speech resynthesis from semantic tokens. Furthermore, we show that the overall UniCATS framework achieves state-of-the-art performance in both speech continuation for zero-shot speaker adaptation and speech editing.
The main contributions of this work are as follows:
\begin{itemize}
    \item We propose a unified context-aware TTS framework called UniCATS to address both speech continuation and editing, which achieves state-of-the-art performance on both the two tasks.
    \item We introduce contextual VQ-diffusion within CTX-txt2vec, enabling the generation of sequence data that seamlessly concatenates with its surrounding context.
    \item We introduce contextual vocoding within CTX-vec2wav to take into consideration the acoustic context when converting the semantic tokens into waveforms.
\end{itemize}

\section{UniCATS}
In this study, we propose a unified context-aware TTS framework called UniCATS, designed to address both speech continuation and editing tasks. UniCATS comprises two components: an acoustic model CTX-txt2vec and a vocoder CTX-vec2wav.
In the following sections, we describe these two components respectively.

\subsection{CTX-txt2vec with Contextual VQ-Diffusion}
CTX-txt2vec employs contextual VQ-diffusion to predict semantic tokens from the input text, enabling it to incorporate the semantic context and maintain seamless concatenation with the surrounding context. We leverage vq-wav2vec tokens as the semantic tokens in this work. 
In this section, we begin by a brief review of VQ-diffusion \cite{vqdiffusion} and then introduce contextual VQ-diffusion. After that, we describe the model architecture, training and inference algorithm of CTX-txt2vec.

\subsubsection{VQ-Diffusion.}

Inspired by diffusion model that has been widely employed in continuous data generation, VQ-diffusion uses a Markovian process for discrete data. Let us consider a data sample consisting of a sequence of discrete indices $\boldsymbol{x}_0=[x_0^{(1)},x_0^{(2)}, ...,x_0^{(l)}]$ where $x_0^{(i)} \in \{1, 2, ..., K\}$. 
During each forward diffusion step, the indices in $\boldsymbol{x}_0$ undergo masking, substitution, or remain unchanged. Following $t$ steps of corruption, the resulting sequence is denoted as $\boldsymbol{x}_t$. For simplicity, we omit the superscript $i$ in the following description. Formally, the equation representing the forward process is
\begin{equation}
q(x_t|x_{t-1}) = \boldsymbol{v}^\top(x_{t})\boldsymbol{Q}_t \boldsymbol{v}(x_{t-1})
\end{equation}
where $\boldsymbol{v}(x_t)\in \mathbb{R}^{(K+1)}$ represents a one-hot vector where $x_t=k$, indicating that only the $k$-th value is 1 while the remaining values are 0. The index value $K+1$ corresponds to the special \texttt{[mask]} token. $\boldsymbol{Q}_t\in \mathbb{R}^{(K+1)\times (K+1)}$ denotes the transition matrix for the $t$-th step. By integrating multiple forward steps, we obtain
\begin{equation}
q(x_t|x_0) = \boldsymbol{v}^\top(x_{t}) \overline{\boldsymbol{Q}}_t \boldsymbol{v}(x_0)
\label{eq:forward}
\end{equation}
where $\overline{\boldsymbol{Q}}_t=\boldsymbol{Q}_t \cdots \boldsymbol{Q}_1$. Applying Bayesian's rule, we have
\begin{equation}
\begin{aligned}
q(x_{t-1}|x_t,x_0) = \frac{q(x_t|x_{t-1},x_0)q(x_{t-1}|x_0)}{q(x_t|x_0)} \\ 
= \frac{
	\left(\boldsymbol{v}^\top(x_{t}) {\boldsymbol{Q}}_t \boldsymbol{v}(x_{t-1})\right) 
	\left(\boldsymbol{v}^\top(x_{t-1}) \overline{\boldsymbol{Q}}_{t-1} \boldsymbol{v}(x_0)\right)}
	{\boldsymbol{v}^\top(x_{t}) \overline{\boldsymbol{Q}}_t \boldsymbol{v}(x_0)}.
\end{aligned}
\end{equation}
The VQ-diffusion model is constructed using a stack of Transformer blocks and is trained to estimate the distribution of $\boldsymbol{x}_0$ from $\boldsymbol{x}_t$ conditioned on $\boldsymbol{y}$, denoted as $p_{\theta}(\tilde{\boldsymbol{x}}_0|\boldsymbol{x}_t, \boldsymbol{y})$. As a result, during the backward process, we can sample $\boldsymbol{x}_{t-1}$ given $\boldsymbol{x}_t$ and $\boldsymbol{y}$ from the following equation
\begin{equation}
p_\theta(\boldsymbol{x}_{t-1}|\boldsymbol{x}_t,\boldsymbol{y}) = \sum_{\tilde{\boldsymbol{x}}_0} q(\boldsymbol{x}_{t-1}|\boldsymbol{x}_t,\tilde{\boldsymbol{x}}_0) p_\theta(\tilde{\boldsymbol{x}}_0|\boldsymbol{x}_t, \boldsymbol{y}).
\label{eq:posterior}
\end{equation}

\subsubsection{Contextual VQ-Diffusion.} 
This study focuses on speech editing and continuation tasks, where the input text serves as the condition $\boldsymbol{y}$, and the semantic tokens to be generated are represented by the data $\boldsymbol{x}_0$. In contrast to the standard VQ-diffusion approach mentioned above, our generation process also takes into account additional context tokens $\boldsymbol{c}^A$ and $\boldsymbol{c}^B$ associated with the data $\boldsymbol{x}_0$. Consequently, we need to model the probability of 
\begin{equation}
p_\theta(\tilde{\boldsymbol{x}}_0|\boldsymbol{x}_t, \boldsymbol{y}, \boldsymbol{c}^A, \boldsymbol{c}^B). 
\end{equation}
To facilitate contextual VQ-diffusion, we propose concatenating the corrupted semantic tokens $\boldsymbol{x}_t$ at diffusion step $t$ with their clean preceding and following context tokens $\boldsymbol{c}^{A}$ and $\boldsymbol{c}^{B}$ in chronological order. This combined sequence, denoted as $[\boldsymbol{c}^{A}, \boldsymbol{x}_t, \boldsymbol{c}^{B}]$, is then fed into the Transformer-based VQ-diffusion model. By doing so, our model can effectively integrate the contextual information using the self-attention layers of the Transformer-based blocks. Similar to Equation \ref{eq:posterior}, we can now calculate the posterior using 
\begin{equation}
\begin{aligned}
& p_\theta(\boldsymbol{x}_{t-1}|\boldsymbol{x}_t,\boldsymbol{y}, \boldsymbol{c}^A, \boldsymbol{c}^B) \\
= & \sum_{\tilde{\boldsymbol{x}}_0} q(\boldsymbol{x}_{t-1}|\boldsymbol{x}_t,\tilde{\boldsymbol{x}}_0) p_\theta(\tilde{\boldsymbol{x}}_0|\boldsymbol{x}_t, \boldsymbol{y}, \boldsymbol{c}^A, \boldsymbol{c}^B).
\end{aligned}
\label{eq:posterior_new}
\end{equation}

\subsubsection{Model Architecture.} 
The architecture of CTX-txt2vec is depicted in Figure \ref{ict2v_overall}, consisting of a text encoder, a duration predictor, a length regulator, and a VQ-diffusion decoder. The sequence of text or phoneme tokens are first encoded by the text encoder, which comprises Transformer blocks, and then employed for duration prediction. Subsequently, the output of the text encoder is expanded based on the corresponding duration values, resulting in the text encoding $\boldsymbol{h}$ that matches the length of semantic tokens. This process follows the idea introduced in FastSpeech 2 \cite{fastspeech2}.

Figure \ref{vqdiff} illustrates the architecture of the VQ-diffusion decoder. The corrupted data $\boldsymbol{x}_t$, resulting from $t$ diffusion steps, is concatenated with its preceding and following context $\boldsymbol{c}^A$ and $\boldsymbol{c}^B$, forming the input sequence $[\boldsymbol{c}^{A}, \boldsymbol{x}_t, \boldsymbol{c}^{B}]$ for the decoder. To distinguish between the data and context, we utilize a binary indicator sequence of the same length as the input. After converting the indicator sequence into embeddings using an embedding table, it is added to the input and then projected and combined with positional encoding. Our VQ-diffusion blocks, based on Transformer, largely follow the architecture in \cite{vqdiffusion}. However, we incorporate the text encoding $\boldsymbol{h}$ differently. Instead of using cross-attention, we directly add $\boldsymbol{h}$ to the output of self-attention layers after applying linear projections. This adjustment is made to accommodate the strict alignment between $\boldsymbol{h}$ and semantic tokens. After passing through $N$ such blocks, the output is layer-normed, projected, and regularized with Softmax to predict the distribution of $p_\theta(\tilde{\boldsymbol{x}}_0|\boldsymbol{x}_t, \boldsymbol{y}, \boldsymbol{c}^A, \boldsymbol{c}^B)$. As the Transformer-based VQ-diffusion decoder generates an output sequence of the same length as its input, only the output segment corresponding to $\boldsymbol{x}_t$ is considered as $\tilde{\boldsymbol{x}}_0$, while the remaining segments are discarded.

\begin{figure*}[t]
  \centering
  \subfigure[Overall architecture and data preparation for training]{
    \includegraphics[page=6,width=0.57\linewidth,trim=0.75cm 0cm 8.2cm 0cm,clip=true]{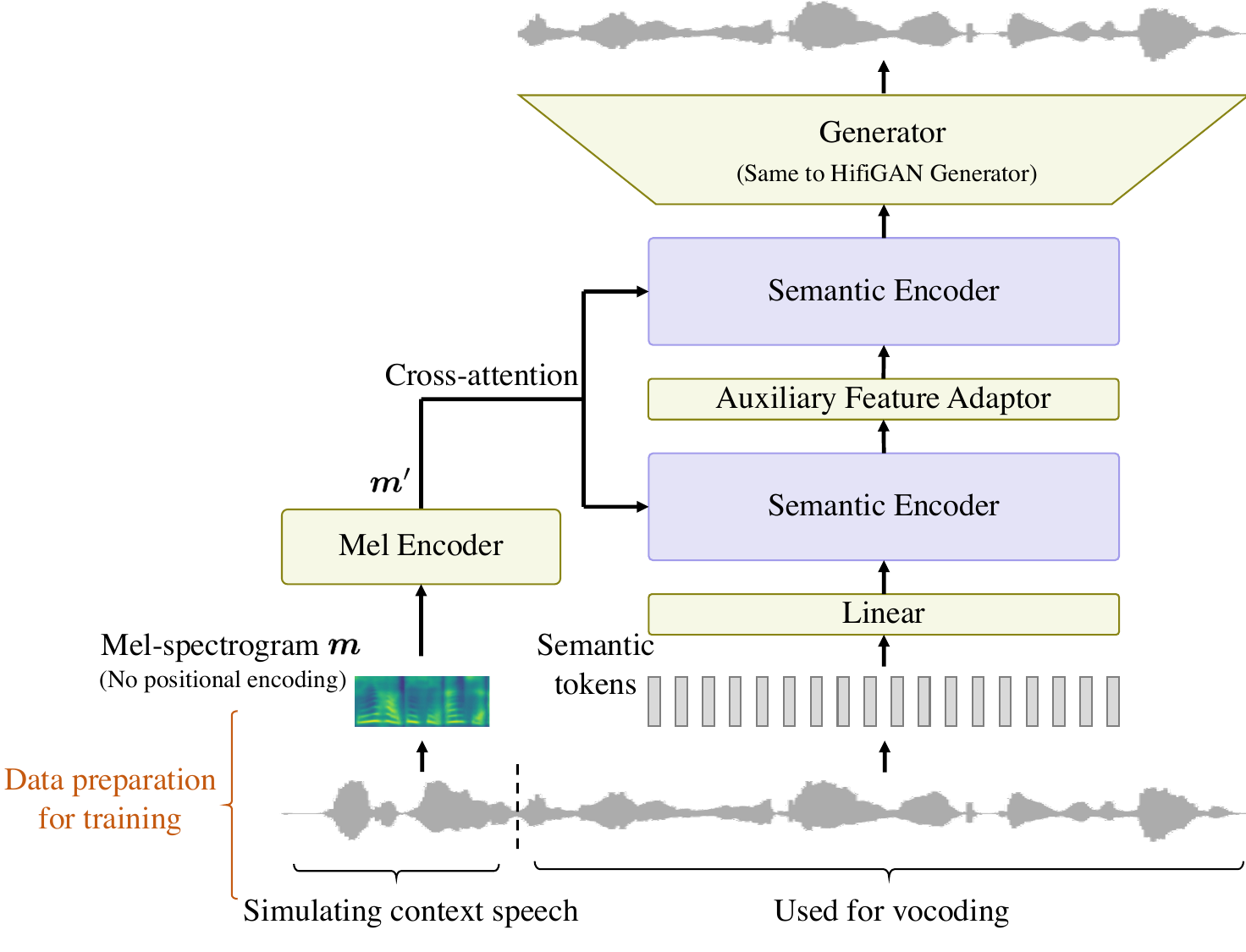}
    \label{icv2w_overall}
  }
  \subfigure[Semantic encoder blocks based on Conformer]{
    \includegraphics[page=7,width=0.35\linewidth,trim=9cm 0cm 9cm 1cm,clip=true]{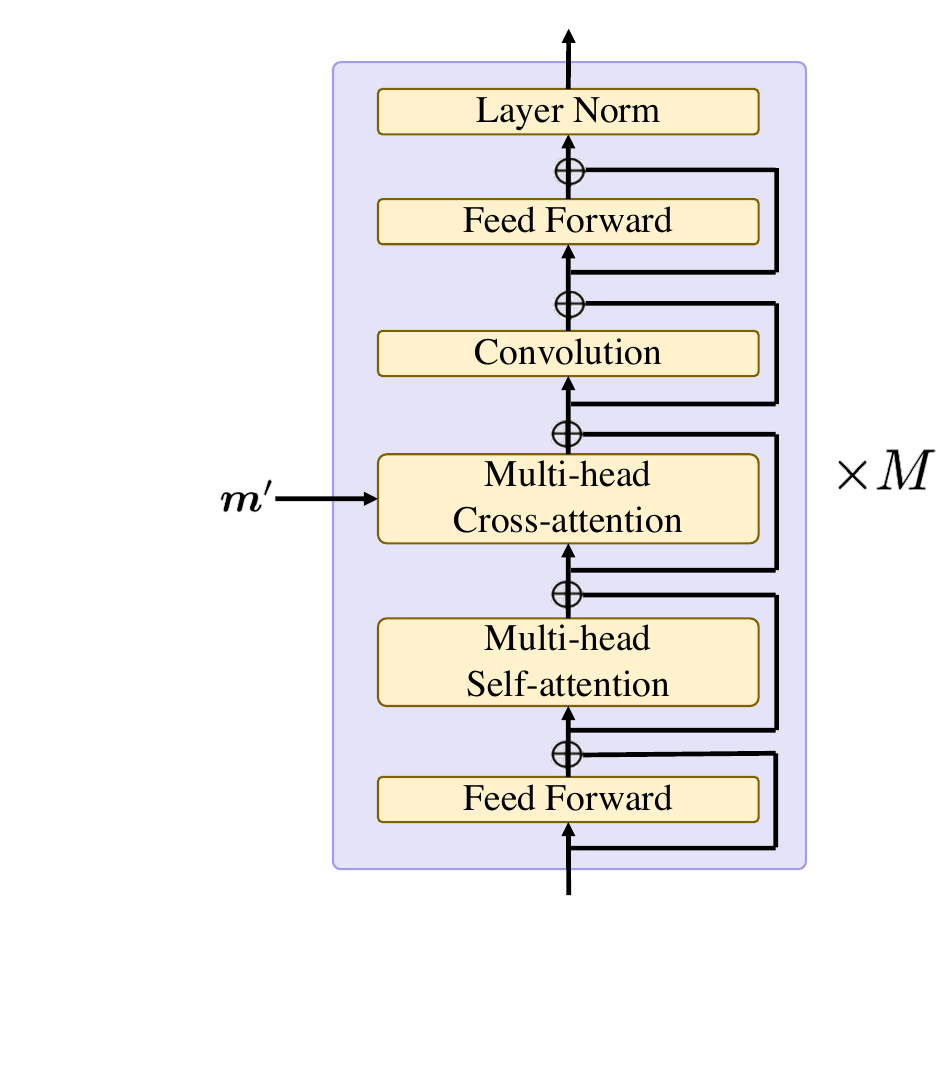}
    \label{conformer}
  }
  \caption{The model architecture of CTX-vec2wav with contextual vocoding.}
  \label{fig:icv2w}
\end{figure*}

\subsubsection{Training Scheme.}

During training, each utterance is randomly utilized in one of three different configurations: with both context A and B, with only context A, or with no context. In the first configuration, the utterance is randomly divided into three segments: context A, $\boldsymbol{x}_0$, and context B. To be specific, we first randomly determine the length of $\boldsymbol{x}_0$, which must be longer than 100 frames yet shorter than the total length of the utterance itself. Next, we randomly determine the starting position of $\boldsymbol{x}_0$. The segments on the left and right sides of $\boldsymbol{x}_0$ are considered as context A and B, respectively.
In the second configuration, we randomly determine the length of context A to be 2-3 seconds. We consider the initial segment of this determined length as context A, while the remaining segment on the right side of context A is assigned as $\boldsymbol{x}_0$.
In the third configuration, the entire utterance is treated as $\boldsymbol{x}_0$ without any context.
The proportion of the three configurations is set to 0.6, 0.3, and 0.1, respectively.

Once the division of the context and the data to be generated is determined, we proceed to corrupt $\boldsymbol{x}_0$ using Equation \ref{eq:forward}, resulting in $\boldsymbol{x}_t$. Subsequently, this corrupted segment, along with its associated context if applicable, is concatenated and utilized as the input for the VQ-diffusion decoder.

The training criterion for CTX-txt2vec, denoted as $\mathcal{L}_{\texttt{CTX-txt2vec}}$, is determined by the weighted summation of the mean square error of duration prediction $\mathcal{L}_{\texttt{duration}}$ and the VQ-diffusion loss $\mathcal{L}_{\texttt{VQ-diffusion}}$ as introduced in \cite{vqdiffusion}, that is 
\begin{equation}
\mathcal{L}_{\texttt{CTX-txt2vec}} = \mathcal{L}_{\texttt{duration}} + \gamma \mathcal{L}_{\texttt{VQ-diffusion}}
\label{eq:t2v_loss}
\end{equation}
where $\gamma$ is a hyper-parameter.

\subsubsection{Inference Algorithm.}

The inference process for speech editing is outlined in Algorithm \ref{alg:t2v}. We first concatenate the phonemes of the speech to be generated, denoted as $\boldsymbol{y}^D$, with the provided context phonemes. This combined sequence is then fed into the text encoder for duration prediction. The predicted duration $\tilde{\boldsymbol{d}}^D$, corresponding to $\boldsymbol{y}^D$, is rescaled using a factor $\alpha$ to maintain a similar speech speed to that of the context.
Then, we iteratively refine the data starting from fully corrupted $\boldsymbol{x}_T$ with its context semantic tokens, following the backward procedure of VQ-diffusion. 
Finally, we obtain the edited semantic tokens $[\boldsymbol{c}^A, \boldsymbol{x}_0, \boldsymbol{c}^B]$.

\begin{algorithm}[tb]
\caption{Inference of CTX-txt2vec for speech editing.}
\label{alg:t2v}
\textbf{Input}: The phonemes, durations and semantic tokens of Context A and B, referred to as $\boldsymbol{y}^A$, $\boldsymbol{y}^B$, $\boldsymbol{d}^A$, $\boldsymbol{d}^B$, $\boldsymbol{c}^A$, $\boldsymbol{c}^B$. The phonemes of speech to be generated $\boldsymbol{y}^D$. \\
\textbf{Parameter}: The number of diffusion steps $T$, fully corrupted tokens $\boldsymbol{x}_T$.\\
\textbf{Output}: Edited semantic tokens.
\begin{algorithmic}[1] 
\STATE $\boldsymbol{y} = [\boldsymbol{y}^A, \boldsymbol{y}^D, \boldsymbol{y}^B]$
\STATE $\boldsymbol{e} = \texttt{TextEncoder}(\boldsymbol{y})$
\STATE $[\tilde{\boldsymbol{d}}^A, \tilde{\boldsymbol{d}}^D, \tilde{\boldsymbol{d}}^B] = \texttt{DurationPredictor}(\boldsymbol{e})$
\STATE $\alpha = (\boldsymbol{d}^A + \boldsymbol{d}^B) / (\tilde{\boldsymbol{d}}^A + \tilde{\boldsymbol{d}}^B)$
\STATE $\boldsymbol{h} = \texttt{LengthRegulator}(\boldsymbol{e}, [\boldsymbol{d}^A, \alpha\tilde{\boldsymbol{d}}^D, \boldsymbol{d}^B])$
\FOR{$t=T, T-1, ..., 1$}
\STATE $p_\theta(\tilde{\boldsymbol{x}}_0|\boldsymbol{x}_t, \boldsymbol{y}, \boldsymbol{c}^A, \boldsymbol{c}^B) = \newline
\hspace*{3em}\texttt{VQDiffusionDecoder}(
[\boldsymbol{c}^A, \boldsymbol{x}_t, \boldsymbol{c}^B], t, \boldsymbol{h})$
\STATE $\boldsymbol{x}_{t-1} \sim p_\theta(\boldsymbol{x}_{t-1}|\boldsymbol{x}_t, \boldsymbol{y}, \boldsymbol{c}^A, \boldsymbol{c}^B)$ \newline
\hspace*{3em} calculated by Equation (\ref{eq:posterior_new})
\ENDFOR
\STATE \textbf{return} $[\boldsymbol{c}^A, \boldsymbol{x}_0, \boldsymbol{c}^B]$
\end{algorithmic}
\end{algorithm}

\subsection{CTX-vec2wav with Contextual Vocoding}

We introduce contextual vocoding within CTX-vec2wav to take into consideration the acoustic context, especially speaker identity, when converting the semantic tokens into waveforms. Consequently, we eliminate the use of speaker embedding and acoustic tokens. In this section, we delve into the architecture of CTX-vec2wav and outline its training scheme.

\subsubsection{Model Architecture.}

The architecture of CTX-vec2wav is illustrated in Figure \ref{icv2w_overall}. The semantic tokens are first projected and encoded through two semantic encoders. Then, the results are passed through convolution and upsampling layers, which are identical to the generator in HifiGAN \cite{hifigan}, to generate the waveforms. An optional auxiliary feature adaptor is set between the two semantic encoders. This module, akin to the variance adaptor in FastSpeech 2, facilitates conditioning the generation on three-dimensional auxiliary features: pitch, energy, and probability of voice (POV) \cite{kaldi_pitch}. Through preliminary experiments, we have observed improvement in audio quality by utilizing this module. As a result, we incorporate it in our model throughout this paper. During training, the model uses ground-truth auxiliary features as conditions and learns to predict them from the output of the first semantic encoder using a projection layer. During inference, the predicted auxiliary features are utilized as conditions.

The literature \cite{audiolm, smantic_resyn} reveals that semantic tokens primarily capture articulation information while lacking sufficient acoustic details, particularly in relation to speaker identity. Therefore, CTX-vec2wav proposes a novel approach of leveraging the mel-spectrogram $\boldsymbol{m}$ to prompt the acoustic contexts, as opposed to conventional methods such as x-vectors \cite{xvector} or acoustic tokens from audio codec models. The semantic encoders in CTX-vec2wav consist of $M$ Conformer-based blocks \cite{conformer}, each of which incorporates an additional cross-attention layer compared to the vanilla Conformer block, enabling the integration of acoustic contexts from the mel-spectrogram. We depict its architecture in Figure \ref{conformer}. Before entering the cross-attention layer, the mel-spectrogram $\boldsymbol{m}$ is encoded by a mel encoder into $\boldsymbol{m}^{\prime}$ using a simple 1D convolution layer in order to integrate consecutive frames.

Note that the mel-spectrogram has no position encoding, resulting in $\boldsymbol{m}^{\prime}$ being a collection of unordered features. This characteristic allows us to utilize mel-spectrograms of varying lengths to prompt acoustic contexts with cross-attention during inference, even though we only utilize a short segment of the mel-spectrogram during training. Increasing the length of the mel-spectrogram has the potential to improve speaker similarity. However, we do not involve this issue in this paper and leave it to be discussed in the future works.

\subsubsection{Training Scheme.}

To effectively utilize speech datasets with inaccurate or absent speaker labels during the training of CTX-vec2wav, we make an assumption that the speaker identity remains consistent within each training utterance. Based on this assumption, we divide each utterance into two segments, as illustrated in Figure \ref{icv2w_overall}. The first segment, which varies randomly in length between 2 to 3 seconds, is utilized for extracting mel-spectrograms and prompting acoustic contexts. The second segment comprises the remaining portion and is used for extracting semantic tokens and performing vocoding.
The training process of CTX-vec2wav follows the same criterion as HifiGAN with an additional L1 loss for auxiliary features prediction. Furthermore, we adopt the multi-task warmup technique proposed in \cite{vqtts}.

\begin{table*}[ht]
\centering
\begin{tabular}{ccc|ccc}
\hline
\textbf{Method}  & \textbf{Feature for Resynthesis}  & \textbf{Speaker Control}  & \textbf{Naturalness MOS} & \textbf{Similarity MOS} & \textbf{SECS}   \\ \hline  \hline
 Ground-truth    & -  & - & 4.91 $\pm$ 0.04 & 4.51 $\pm$ 0.08 & 0.837 \\
 Encodec        & Acoustic token  &  - & 4.39 $\pm$ 0.07 & 4.00 $\pm$ 0.08 & 0.829 \\ \hline
 HifiGAN     & Semantic token  & X-vector & 4.30 $\pm$ 0.08  & 3.96 $\pm$ 0.08 &  0.776 \\
AudioLM      & Semantic token & AR continuation & 3.99 $\pm$ 0.07 & 3.96 $\pm$ 0.08 & 0.801  \\
CTX-vec2wav & Semantic token & Contextual vocoding & \textbf{4.75 $\pm$ 0.06} & \textbf{4.50 $\pm$ 0.07} & \textbf{0.845} \\ \hline 
\end{tabular}
\caption{The performance of speech resynthesis from semantic tokens.}
\label{tab:resyn}
\end{table*}

\subsection{Unified Framework for Context-Aware TTS}

UniCATS prompts semantic and acoustic contexts through their respective semantic tokens and mel-spectrograms. Following Algorithm \ref{alg:t2v}, the edited semantic tokens $[\boldsymbol{c}^A, \boldsymbol{x}_0, \boldsymbol{c}^B]$ are obtained. These tokens are then vocoded into waveforms, with the speaker information indicated by the mel-spectrogram of the contexts $[\boldsymbol{m}^A, \boldsymbol{m}^B]$.

Since the only distinction between speech continuation and editing lies in the presence or absence of context B, all the aforementioned algorithms for speech editing can be readily generalized to speech continuation by excluding context B. Consequently, UniCATS demonstrates the capability to handle both the two context-aware TTS tasks.


\section{Experiments and Results}

\subsection{Dataset}

LibriTTS is a multi-speaker transcribed English speech dataset. Its training set consists of approximately 580 hours of speech data from 2,306 speakers. For evaluation purposes, we exclude 500 utterances from the official LibriTTS training set, which will serve as one of our test sets referred to as ``test set A''. Test set A comprises 369 speakers out of the 2,306 training speakers.
In addition, we utilize 500 utterances from the ``test-clean'' set of LibriTTS, designated as ``test set B'', to assess the zero-shot adaptation capability for new and unseen speakers. Test set B contains 37 speakers. Each speaker in test sets A and B is associated with a brief speech prompt lasting approximately 3 seconds. The utterance list for both test sets A and B, along with their corresponding prompts, is available on our demo page. 
Lastly, for evaluating speech editing, we employ the same test set as utilized in \cite{retrievertts}. The utterances for this evaluation are also derived from the ``test-clean'' set of LibriTTS and are denoted as ``test set C''.

 \subsection{Training Setup}

In CTX-txt2vec, the text encoder consists of 6 layers of Transformer blocks. The VQ-diffusion decoder employs $N=12$ Transformer-based blocks with attention layers comprising 8 heads and a dimension of 512. In Equation \ref{eq:t2v_loss}, the value of $\gamma$ is set to 1. The semantic tokens are extracted using a pretrained kmeans-based vq-wav2vec model\footnote{https://github.com/facebookresearch/fairseq/tree/main/examples\\/wav2vec}. CTX-txt2vec is trained for 50 epochs using an AdamW \cite{adamw} optimizer with a weight decay of $4.5 \times 10^{-2}$. The number of diffusion steps is set to $T=100$.
In CTX-vec2wav, both semantic encoders consist of $M=2$ Conformer-based blocks. The attention layers within these blocks have 2 heads and a dimension of 184. The mel encoder employs a 1D convolution with a kernel size of 5 and an output channel of 184. CTX-vec2wav is trained using an Adam \cite{adam} optimizer for 800k steps. The initial learning rate is set to $2 \times 10^{-4}$ and is halved every 200k steps.

\begin{table*}[ht]
\centering
\begin{tabular}{c|ccc|ccc}
\hline
\multirow{2}{*}{\textbf{Method}} & \multicolumn{2}{c}{\textbf{\hspace{4em}Seen Speakers}} & & \multicolumn{3}{c}{\textbf{Unseen Speakers}}\\ \cline{2-7} 
 & \textbf{Naturalness MOS} & \textbf{Similarity MOS} & \textbf{SECS}  & \textbf{Naturalness MOS} & \textbf{Similarity MOS} & \textbf{SECS}   \\ \hline  \hline
 Ground-truth  & 4.89 $\pm$ 0.04 & 4.54 $\pm$ 0.08 & 0.833 & 4.91 $\pm$ 0.04 &  4.50 $\pm$ 0.08  &  0.837 \\ \hline
 FastSpeech 2  & 3.81 $\pm$ 0.08 & 3.94 $\pm$ 0.06 & 0.820 & 3.65 $\pm$ 0.08 &  3.65 $\pm$ 0.07  &  0.770   \\
VALL-E         & 4.23 $\pm$ 0.07 & 3.98 $\pm$ 0.06 & 0.796 & 4.17 $\pm$ 0.09 &  3.83 $\pm$ 0.07  &   0.786    \\
UniCATS        & \textbf{4.54 $\pm$ 0.07} & \textbf{4.56 $\pm$ 0.07} & \textbf{0.831} & \textbf{4.43 $\pm$ 0.08} &  \textbf{4.25 $\pm$ 0.08} &    \textbf{0.836} \\ \hline 
\end{tabular}
\caption{The performance of zero-shot speaker adaptative text-to-speech.}
\label{tab:adapt}
\end{table*}

\subsection{Speech Resynthesis from Semantic Tokens}

We begin by examining the performance of CTX-vec2wav in speech resynthesis from semantic tokens on test set B. Two common methods for vocoding semantic tokens, namely HifiGAN and AudioLM, are used as baselines in our evaluation. We utilize the open-source implementation of AudioLM\footnote{https://github.com/lucidrains/audiolm-pytorch}, as we do not have access to its official internal implementation.
Each speaker in the test set is associated with a brief speech prompt that indicates the speaker's identity. In HifiGAN vocoding, we employ a pretrained speaker-verification model\footnote{https://huggingface.co/speechbrain/spkrec-ecapa-voxceleb} to extract x-vectors from the prompts, enabling us to control the speaker information, following the idea presented in \cite{smantic_resyn}. In AudioLM decoding, we use acoustic tokens from the prompts for AR continuation.
In CTX-vec2wav, as previously discussed, we use the mel-spectrogram of the prompt to control the speaker identity by contextual vocoding. All these models are trained to resynthesize speech from the same semantic tokens extracted by vq-wav2vec. We also evaluate the official Encodec model\footnote{https://github.com/facebookresearch/encodec} for resynthesizing speech from the acoustic tokens, which is theoretically an easier task. 

We evaluate the generated results using MOS listening tests, where 15 listeners rate the presented utterances on a scale of 1 to 5 in terms of naturalness and speaker similarity to the prompt. Additionally, we compute the Speaker Encoder Cosine Similarity (SECS) \cite{yourtts} as an auxiliary metric to assess speaker similarity. The SECS scores are calculated using the speaker encoder in Resemblyzer\footnote{https://github.com/resemble-ai/Resemblyzer}.

The results are shown in Table \ref{tab:resyn}.
Our proposed CTX-vec2wav demonstrates the best performance in speech resynthesis from semantic tokens in terms of both naturalness and speaker similarity. 
In contrast, when vocoding semantic tokens with HifiGAN, we observe the lowest SECS score. This can be attributed to the information compression inherent in x-vectors as a bottleneck feature of the speaker-verification model. Although x-vectors effectively distinguish between speakers, they are not ideal for accurately reconstructing the speaker's voice.
Remarkably, CTX-vec2wav even outperforms Encodec in subjective evaluations and achieves a SECS score comparable to the ground truth.
We notice that Encodec resynthesis occasionally introduces artifacts that negatively impact the subjective scores.

\subsection{Speech Continuation for Zero-Shot Speaker Adaptation}

In this section, we assess the performance of UniCATS in zero-shot speaker adaptation with speech continuation. We utilize test sets A and B for evaluating seen and unseen speakers respectively. Our baselines include x-vector-based multi-speaker TTS model FastSpeech 2 from ESPnet toolkit \cite{espnet} and the state-of-the-art zero-shot speaker adaptive TTS model VALL-E.
As the official implementation of VALL-E is not publicly available, we employ the open-source VALL-E model\footnote{https://github.com/lifeiteng/vall-e} trained on LibriTTS for our evaluation.
To evaluate the generated results, we conduct MOS listening tests following the same methodology as described in the previous section. 15 listeners are asked to rate the presented utterances on a scale of 1 to 5, considering naturalness and speaker similarity to the prompt. Similarly, we introduce SECS as another metric to assess speaker similarity.

We demonstrate the results in Table \ref{tab:adapt}. For seen speakers,  UniCATS achieves a much better naturalness compared with the FastSpeech 2 and VALL-E baselines. FastSpeech 2 has a relatively limited naturalness due to the use of mel-spectrogram, while VALL-E's performance is limited by the performance of Encodec. The speaker similarity of UniCATS is close to the ground-truth and outperforms the two baselines in both subjective and objective evaluations. For unseen speakers, UniCATS also achieves the best performance in terms of both naturalness and speaker similarity. However, all systems perform slightly worse for unseen speakers than for seen speakers in the subjective scores.

The results are reported in Table \ref{tab:adapt}. In the case of seen speakers, UniCATS achieves significantly better naturalness compared to the FastSpeech 2 and VALL-E baselines. FastSpeech 2's naturalness is relatively limited due to its reliance on mel-spectrograms, while VALL-E's performance is constrained by the capabilities of Encodec.
UniCATS also achieves speaker similarity scores that is quite close to the ground truth, outperforming both baselines in both subjective and objective metrics. All systems demonstrate slightly diminished subjective scores for unseen speakers when compared to seen speakers.

It is worth noting that MOS scores for naturalness and speaker similarity achieved by UniCATS are even higher than those of Encodec resynthesis, indicating that our model breaks the upper bound of a series of other works that uses acoustic tokens.

\subsection{Speech Editing}

We utilize test set C to evaluate speech editing, where each utterance is divided into three segments: context A, the segment $\boldsymbol{x}$ to be generated, and context B. This division allows us to simulate speech editing and compare the generated results with the ground truth. To evaluate short and long segment editing separately, we employ two different segment division approaches. For short editing, $\boldsymbol{x}$ consists of randomly chosen 1 to 3 words. For long editing, $\boldsymbol{x}$ contains as many words as possible while remaining within a 2-second duration. In our experiments, we compare UniCATS with the state-of-the-art speech editing model RetrieverTTS \cite{retrievertts}. In the MOS listening test, participants are requested to rate the naturalness of the generated segments $\boldsymbol{x}$ and their contextual coherence.

\begin{table}[ht]
\centering
\begin{tabular}{c|cc}
\hline
\textbf{Method} & \textbf{MOS@short} & \textbf{MOS@long}  \\ \hline  \hline
 Ground-truth  & 4.77 $\pm$ 0.06    & 4.90 $\pm$ 0.04  \\ \hline
 RetrieverTTS  &  4.43 $\pm$ 0.08   & 4.37 $\pm$ 0.08  \\
UniCATS        & \textbf{4.62 $\pm$ 0.06}  & \textbf{4.63 $\pm$ 0.06}  \\ \hline 
\end{tabular}
\caption{The performance of speech editing.}
\label{tab:edit}
\end{table}

The results in Table \ref{tab:edit} demonstrate that UniCATS outperforms RetrieverTTS in both scenarios. Moreover, as the length of the editing segment increases, the performance of RetrieverTTS declines. Conversely, UniCATS exhibits consistent performance across varying segment lengths.

\section{Conclusion}
In this work, we propose a unified context-aware TTS framework called UniCATS, designed to handle both speech continuation and editing tasks. UniCATS eliminates the use of acoustic tokens and speaker embeddings. Instead, it utilizes contextual VQ-diffusion and vocoding in CTX-txt2vec and CTX-vec2wav respectively for incorporating both semantic and acoustic context information. 
Our experiments conducted on the LibriTTS dataset demonstrate that CTX-vec2wav outperforms HifiGAN and AudioLM in terms of speech resynthesis from semantic tokens. Furthermore, we show that the overall UniCATS framework achieves state-of-the-art performance in both speech continuation for zero-shot speaker adaptation and speech editing.

\appendix

\section{Word Error Rate of Resynthesis}

In this work, we opt to utilize vq-wav2vec tokens that retain richer prosody information than other semantic tokens such as HuBERT. In this section, we dive into evaluating the word error rate (WER) of resynthesis from HuBERT and vq-wav2vec. To this end, we train two CTX-vec2wav models with the two types of semantic tokens respectively. Then we resynthesize the speech in test set B from HuBERT and vq-wav2vec tokens and calculate their WERs respectively with Whisper \cite{whisper}, a well-known automatic speech recognition (ASR) model. The results are shown in Table \ref{tab:asr}.

\begin{table}[ht]
\centering
\begin{tabular}{ccc}
\hline
\textbf{Ground-truth} & \textbf{HuBERT} & \textbf{Vq-wav2vec}  \\ \hline 
1.83        &  2.73  & 3.64 \\ \hline 
\end{tabular}
\caption{The word error rate of speech reconstruction from different semantic tokens.}
\label{tab:asr}
\end{table}

We can see that vq-wav2vec has a higher WER than HuBERT, although it contains richer prosody information. Therefore, none of the two types of semantic tokens achieve the best of both worlds. A better speech representation is still worthy exploring in the future work.

\bibliography{aaai24}

\end{document}